\documentclass[12pt,preprint]{iopart}
\usepackage{amssymb}
\usepackage{graphicx}

\begin{document}

\title[FPT of a generic Brownian particle in a time-inhomogeneous setup]{First-passage-time statistics of a Brownian particle driven by an arbitrary unidimensional potential with a superimposed exponential time-dependent drift}

\author{Eugenio Urdapilleta}
\address{Divisi\'on de F\'isica Estad\'istica e Interdisciplinaria, Centro At\'omico Bariloche, Av. E. Bustillo Km 9.500, S. C. de Bariloche (8400), R\'io Negro, Argentina}
\ead{eugenio.urdapilleta@cab.cnea.gov.ar}
\begin{abstract}
In one-dimensional systems, the dynamics of a Brownian particle
are governed by the force derived from a potential as well as by
diffusion properties. In this work, we obtain the
first-passage-time statistics of a Brownian particle driven by an
arbitrary potential with an exponential temporally decaying
superimposed field up to a prescribed threshold. The general
system analyzed here describes the sub-threshold signal
integration of integrate-and-fire neuron models, of any kind,
supplemented by an adaptation-like current, whereas the
first-passage-time corresponds to the declaration of a spike.
Following our previous studies, we base our analysis on the
backward Fokker-Planck equation and study the survival probability
and the first-passage-time density function in the space of the
initial condition. By proposing a series solution we obtain a
system of recurrence equations, which given the specific structure
of the exponential time-dependent drift, easily admit a simpler
Laplace representation. Naturally, the present general derivation
agrees with the explicit solution we found previously for the
Wiener process in (2012 {\it J. Phys. A: Math. Theor.} {\bf 45}
185001). However, to demonstrate the generality of the approach,
we further explicitly evaluate the first-passage-time statistics
of the underlying Ornstein-Uhlenbeck process. To test the validity
of the series solution, we extensively compare theoretical
expressions with the data obtained from numerical simulations in
different regimes. As shown, agreement is precise whenever the
series is truncated at an appropriate order. Beyond the fact that
both Wiener and Ornstein-Uhlenbeck processes have a direct
interpretation in the context of neuronal models, given their
ubiquity in different fields, our present results will be of
interest in other settings where an additive state-independent
temporal relaxation process is being developed as the particle
diffuses.
\end{abstract}

\maketitle

\section{Introduction}\label{intro}
\indent From the first attempts in the nineteenth century to
account for the effects of randomness in nature, fluctuations have
been recognized as a fundamental component in the description of
several physical, chemical and biological systems \cite{Gardiner,
Nelson, VanKampen}. A stochastic description can be formulated in
continuous or discrete time for systems whose phase space is
continuous, or embedded in a lattice. For many of these systems,
Brownian motion and diffusion models have proven to be key
conceptualizations, mathematically accessible and useful for
characterizing their statistical behavior \cite{Gardiner, Nelson,
VanKampen, Risken, Ricciardi, HanggiMarchesoni2005}.

\indent Random walks and diffusion processes evolve in certain
regions of their phase space according to the equations or rules
governing their dynamics. Complementary to dynamics, boundary
conditions shape the probability distribution of the system being
in a given state at certain time. In this sense, different
scenarios are prescribed by different combinations of boundary
conditions \cite{Gardiner, VanKampen, Risken}. Among these
scenarios, first-passage-time (FPT) problems -also referred to as
exit or escape times- represent a wide class of situations
\cite{Redner}, where the random variable of interest is the time
at which certain state conditions are met first. The statistics of
this and other related variables, such as survival probability,
are relevant in many fields and contexts \cite{Redner,
Metzler_etal, Grebenkov2015, Ryabov_etal2015}. Examples abound in
different disciplines: diffusion-influenced reactions
\cite{Tachiya1979, SanoTachiya1979, Szabo_etal1980,
Hanggi_etal1990, Krapivsky_etal1994, Krapivsky2012}, the
channel-assisted membrane transport of metabolites, ions, or
polymers \cite{Eisenberg_etal1995, Berezhkovskii_etal2002,
GoychukHanggi2002, Cohen_etal2012} and, in general, the variety of
processes orchestrating intracellular transport
\cite{BressloffNewby2013}, force-induced unbinding processes and
the reconstruction of potential functions in experiments of
single-molecule force spectroscopy \cite{Balsera_etal1997,
HummerSzabo2003, Dudko_etal2006, Dudko_etal2008, FokChou2010}, the
random search and first encounters of mobile or immobile targets
by a variety of agents, from foraging animals to proteins
\cite{Benichou_etal2011}, including transport-limited reactions in
active media \cite{Loverdo_etal2008}, the absorption of particles
in restricted geometries \cite{MeersonRedner2014,
RednerMeerson2014} and, generically, geometry-controlled kinetics
\cite{BenichouVoituriez2014}, escape from confined domains through
narrow pores \cite{BenichouVoituriez2008, Pillay_etal2010,
Cheviakov_etal2010} as well as through large windows
\cite{Rupprecht_etal2015}, the granular segregation of binary
mixtures \cite{FarkasFulop2001}, the formation of loops in nucleic
acids and the cyclization of polymers \cite{Sokolov2003,
Guerin_etal2012}, and finally, the nucleation and stochastic
self-assembly of monomers \cite{Yvinec_etal2012}, to name a few.

\indent A particularly interesting FPT problem arises in the
context of spiking neurons. In general, a neuron accumulates input
currents up to a point at which its dynamics become independent of
the inputs, and then generates a large excursion in its phase
space in a relatively short period of time, producing a kind of
stereotyped event called an {\it action potential} or a {\it
spike} \cite{Koch, GerstnerKistler}. To a large extent, these
spikes form the basis of neuronal communication. Simplified models
of spike generation take advantage of the large reproducibility of
action potentials by splitting the dynamics in two phases: the
phase where inputs and internal processes drive the neuron, i.e.
the regime of sub-threshold integration, and the phase where the
spike is manifested itself, which is not represented in detail but
prescribed by a fire-and-reset rule \cite{Koch, GerstnerKistler}.
These {\it integrate-and-fire} (IF) models have different versions
according to the processes included during the sub-threshold
integration \cite{Koch, GerstnerKistler, Fourcaud-Trocme2003}.
However, all of them produce spikes by declaration once the
voltage reaches a certain threshold for the first time. Then, the
voltage is reset to a lower potential, which is the initial
condition for the sub-threshold integration of the following
spike. Because of how spikes are defined in these models, the
relationship to FPT problems is straightforward. On the other
hand, the reproducibility of {\it spike times} is not as precise
as expected from deterministic systems \cite{Koch,
GerstnerKistler}. This randomness is accounted for by different
stochastic phenomena (ionic gating, neuro-transmitter release,
etc), which are normally incorporated in these models by adding
noise -in general, Gaussian white or colored noise. Overall, the
stochastic dynamics of the sub-threshold integration of IF models
is equivalent to the motion of a Brownian particle (where
different versions correspond to different potentials), whereas
the presence of the threshold for declaring spikes sets the
definition of the FPT \cite{Ricciardi, Koch, GerstnerKistler,
Gerstein1964, Tuckwell, Burkitt2006, SacerdoteGiraudo2013}.

\indent One of the processes that greatly affects spike production
is the presence of adaptation currents. They are putatively
responsible for the widespread observed phenomenon of spike
frequency adaptation, where a neuron's discharge rate decreases in
response to a step stimulus \cite{Koch, GerstnerKistler}. The
simplest version of adaptive IF models incorporates an additive
current that decreases exponentially in time during sub-threshold
evolution, and enables history-dependent behavior in the
spike-and-reset rule \cite{Treves1993, BendaHerz2003,
Urdapilleta2011b, SchwalgerLindner2013}. Therefore, the first step
towards developing a full understanding of the effect of
spike-triggered adaptation currents on interspike interval
statistics is to analyze the FPT problem of a Brownian motion
-given by the associated IF model under analysis- with a
superimposed exponentially decaying additive drift. This problem
was partially covered by Lindner in 2004, in which he derived
general expressions for the corrections to the moments of FPT
distribution in a general time-dependent case, and assessed some
particular explicit solutions to the first order moments in the
case of exponential temporal driving \cite{Lindner2004}. To
address the full FPT statistics of the temporally inhomogeneous
problems that exponential time-dependent drift poses, we
previously analyzed the survival probability and the FPT
distribution based on the backward Fokker-Planck (FP) equation,
and inductively solved the system of infinite recurrence equations
that results from the proposal of a series solution. This approach
was derived for the simple case of a Wiener process, corresponding
to the {\it perfect} IF model in the spiking neurons framework
\cite{Urdapilleta2011a, Urdapilleta2012}. Here, we aim to extend
these results to a system where the main driving force derives
from an arbitrary potential. Furthermore, we were able to obtain
the explicit solution to the Ornstein-Uhlenbeck process, which
corresponds to the {\it leaky} IF model in the context of spiking
neurons. Traditionally, this is considered to be the minimal
biologically realistic IF model \cite{Tuckwell}, obtained as a
diffusion approximation of the Stein model \cite{Stein1965}.

\section{First-passage-time in an arbitrary potential: Theoretical framework}
\subsection{The homogeneous system}\label{unperturbed}
\indent First, the methodology we use later to study the
first-passage-time problem in a time-inhomogeneous setup is
reviewed in a classical context: a particle driven exclusively by
an arbitrary unidimensional potential \cite{Risken, Gardiner}.
Even when different approaches can be taken, we focus on the
analysis of the survival probability via the backward
Fokker-Planck equation. Under this framework, the posterior
analysis of a superimposed exponential time-dependent drift has
proven to be mathematically tractable \cite{Urdapilleta2011a,
Urdapilleta2012}.\\
\indent We consider a continuous-time random walk representing the
movement of an overdamped particle, whose position $x(t)$ evolves
according to the following equation:

\begin{equation}\label{eq1}
   \frac{\rmd x}{\rmd t} = -\frac{\rmd U(x)}{\rmd x} + \sqrt{2D}~\xi(t).
\end{equation}

\indent In Eq.~(\ref{eq1}), the conservative force driving the
particle is written as the negative of the derivative of a generic
position-dependent potential $U(x)$, whereas random forces are
represented by an additive Gaussian white noise $\xi(t)$, defined
by $\langle \xi(t) \rangle = 0$ and $\langle \xi(t) \xi(t')
\rangle = \delta(t-t')$. The particle is initially set at position
$x_0$ and the first time it reaches the level $x_{\rm thr} > x_0$,
the dynamics are no longer analyzed (or reset) and a
first-passage-time (FPT) event is declared.\\
\indent For this system, the evolution of the transition
probability density function of a particle being at position $x$
at time $t$ -given that it will be at position $t'$ at a posterior
time $t'$ ($t' > t$)- is governed by the backward Fokker-Planck
(FP) equation \cite{Risken, Gardiner, Urdapilleta2011a}, which
reads

\begin{equation}\label{eq2}
   \frac{\partial P(x',t'|x,t)}{\partial t} = U'(x)~\frac{\partial
   P(x',t'|x,t)}{\partial x} - D \frac{\partial^2 P(x',t'|x,t)}
   {\partial x^2},
\end{equation}

\noindent where $U'(x)$ indicates the derivative of the potential
with respect to $x$. Supplementing Eq.~(\ref{eq2}), initial and
boundary conditions are set in accordance to the survival domain,
$x'<x_{\rm thr}$. Explicitly,

\begin{eqnarray}
   \label{eq3}
   P(x',t'|x,t=t') &=& \cases{1 ~~{\rm for}~x<x_{\rm thr},\\
   0~~{\rm for}~x\geq x_{\rm thr},}\\
   \label{eq4}
   P(x',t'|x=x_{\rm thr},t) &=& 0.
\end{eqnarray}

\indent Within the backward FP formalism, the survival probability
at time $t'$ for a particle released at position $x$ at time $t$,

\begin{equation}\label{eq5}
   F(t'|x,t) = \int_{-\infty}^{x_{\rm thr}} P(x',t'|x,t) ~{\rmd
   x'},
\end{equation}

\noindent is obtained simply by integration of both sides of
Eq.~(\ref{eq2}) in $x'$ from $-\infty$ to $x_{\rm thr}$. In terms
of the auxiliary variable $\tau = t'-t$, the corresponding
equation results

\begin{equation}\label{eq6}
   \frac{\partial F(x,\tau)}{\partial \tau} = - U'(x)~\frac{\partial
   F(x,\tau)}{\partial x} + D \frac{\partial^2 F(x,\tau)}
   {\partial x^2},
\end{equation}

\noindent whereas initial and boundary conditions read

\begin{eqnarray}
   \label{eq7}
   F(x,\tau=0) &=& \cases{1 ~~{\rm for}~x<x_{\rm thr},\\
   0~~{\rm for}~x\geq x_{\rm thr},}\\
   \label{eq8}
   F(x=x_{\rm thr},\tau) &=& 0.
\end{eqnarray}

\indent Once the survival probability, whose evolution is given by
Eqs.~(\ref{eq6})-(\ref{eq8}), is known, the FPT density function
can be immediately evaluated. By construction, $F(x,\tau)$ is the
probability that -given that the particle was released at position
$x$- it remains alive or not absorbed at time $\tau$ within the
survival domain. Equivalently, $F(x,\tau)$ represents the
probability that the FPT is posterior to $\tau$: ${\rm
Prob}(T>\tau) = F(x,\tau)$, where $T$ is the FPT random variable.
Since we are focusing on the potentials that warranty level
crossing (which can be set, for example, with a small positive
drift), then $F(x,\tau)$ can be directly related to the cumulative
distribution function of the FPT, $\Phi(\tau)$, through the
relationship: $\Phi(\tau) = {\rm Prob}(T \leq \tau) = 1 -
F(x,\tau)$. Therefore, the density function of the FPT,
$\phi(\tau)$, is given by

\begin{equation}\label{eq9}
   \phi(\tau) = \frac{\rmd \Phi(\tau)}{\rmd \tau} = - \frac{\partial
   F(x,\tau)}{\partial \tau},
\end{equation}

\noindent where it should be noted that once the system defined by
Eqs.~(\ref{eq6})-(\ref{eq8}) has been solved, the initial backward
position of the particle $x$ remains as a parameter and can be
removed from the notation.

\subsection{An additive time-dependent exponential drift}

\indent In this work, we study the influence on survival
probability and FPT statistics of a superimposed exponential
time-dependent drift in a system driven by the potential $U(x)$.
This system is described by the Langevin equation

\begin{equation}\label{eq10}
   \frac{\rmd x}{\rmd t} = - U'(x) + \frac{\epsilon}
   {\tau_{\rmd}}~\rme^{-(t-t_0)/\tau_{\rmd}} + \sqrt{2D}~\xi(t),
\end{equation}

\noindent where $\epsilon$ and $\tau_{\rmd}$ characterize the
strength and time constant  of the exponential driving, and
$t_{0}$ refers to the initial time of the experimental setting.\\
\indent The backward FP equation is similar to Eq.~(\ref{eq2}),
except for an additional term in the drift coefficient

\begin{equation}\label{eq11}
   \fl \hspace{0.5cm}
   \frac{\partial P(x',t'|x,t)}{\partial t} = \left[ U'(x)
    - \frac{\epsilon}{\tau_{\rmd}}~\rme^{-(t-t_0)/\tau_{\rmd}} \right]~
    ~\frac{\partial P(x',t'|x,t)}{\partial x} - D \frac{\partial^2
    P(x',t'|x,t)}{\partial x^2}.
\end{equation}

\indent Proceeding as in the previous subsection, the survival
probability $F(x,\tau; t')$, in terms of the auxiliary variable
$\tau = t'-t$, reads

\begin{equation}\label{eq12}
   \fl \hspace{0.5cm}
   \frac{\partial F(x,\tau;t')}{\partial \tau} = \left[ - U'(x)
   + \frac{\epsilon}{\tau_{\rmd}}~\rme^{-(t'-t_0)/\tau_{\rmd}}~\rme^{\tau/\tau_{\rmd}}
   \right]~\frac{\partial F(x,\tau;t')}{\partial x} + D \frac{\partial^2 F(x,\tau;t')}
   {\partial x^2},
\end{equation}

\noindent whereas the initial and boundary conditions are similar
to those given by Eqs.~(\ref{eq7}) and (\ref{eq8}). Note that here
the term ``initial'' refers to the situation $\tau = 0$, and not
to the initial time of the experimental setting, $t_0$.\\
\indent Following our previous studies \cite{Urdapilleta2011a,
Urdapilleta2012}, we propose a series expansion in powers of
$\epsilon$ for $F(x,\tau;t')$,

\begin{equation}\label{eq13}
   \fl \hspace{0.5cm}
   F(x,\tau;t') = F_{0}(x,\tau;t') + \epsilon~F_{1}(x,\tau;t') +
   \epsilon^{2}~F_{2}(x,\tau;t') + \dots =
   \sum_{n=0}^{\infty}\epsilon^{n}~F_{n}(x,\tau;t').
\end{equation}

\indent With this assumption, Eq.~(\ref{eq12}) reads

\begin{eqnarray}\label{eq14}
\fl \hspace{1.5cm}
   \left[ \frac{\partial F_{0}}{\partial \tau} + U'(x) \frac{\partial
   F_{0}}{\partial x} - D \frac{\partial^{2} F_{0}}{\partial x^{2}}
   \right] \nonumber\\
\fl \hspace{1.5cm}
   + \sum_{n=1}^{\infty} \epsilon^{n}~\left[\frac{\partial F_{n}}
   {\partial \tau} + U'(x) \frac{\partial F_{n}}{\partial x} -
   \frac{1}{\tau_{\rmd}}~\rme^{-(t'-t_{0})/\tau_{\rmd}}~\rme^{\tau/\tau_{\rmd}}
   ~\frac{\partial F_{n-1}}{\partial x} - D \frac{\partial^{2} F_{n}}{\partial
   x^{2}}\right] = 0.
\end{eqnarray}

\indent Given the arbitrariness of $\epsilon$, each term between
brackets should be identically $0$, splitting Eq.~(\ref{eq14})
into an infinite system of coupled equations with a recursive
structure,

\begin{eqnarray}
   \label{eq15}
   \fl \hspace{1.5cm}
   \frac{\partial F_{0}}{\partial \tau} + U'(x)~\frac{\partial F_{0}}
   {\partial x} - D~\frac{\partial^{2} F_{0}}{\partial x^{2}} &=& 0, \\
   \label{eq16}
   \fl \hspace{1.5cm}
   \frac{\partial F_{n}}{\partial \tau} + U'(x)~\frac{\partial F_{n}}
   {\partial x} - D~\frac{\partial^{2} F_{n}}{\partial x^{2}}
   &=& \frac{1}{\tau_{\rmd}}~\rme^{-(t'-t_{0})/\tau_{\rmd}}~\rme^{\tau/\tau_{\rmd}}
   ~\frac{\partial F_{n-1}}{\partial x}, ~~{\rm for}~~n \geq 1.
\end{eqnarray}

\indent For the same reason, the non-homogeneous initial
condition, $F(x,\tau=0;t') = 1$ for $x < x_{\rm thr}$, has to be
imposed on the zeroth-order function $F_{0}(x,\tau=0;t')$,
yielding

\begin{eqnarray}\label{eq17}
   F_{0}(x,\tau=0;t') &=& \cases{1 ~~{\rm if}~
   x<x_{\rm thr},\\
   0 ~~{\rm if}~x\geq x_{\rm thr},}\\
   \label{eq18}
   F_{n}(x,\tau=0;t') &=& 0 ~~{\rm for}~ n\geq 1.
\end{eqnarray}

\indent Without specificity of the order, the boundary condition
reads $F_{n}(x=x_{\rm thr},\tau;t') = 0$ for all $n$.\\
\indent The preceding system of recursive \textit{partial}
equations, Eqs.~(\ref{eq15}) and~(\ref{eq16}), can be
Laplace-transformed to an equivalent system of \textit{ordinary}
equations,

\begin{eqnarray}
   \label{eq19}
   \fl \hspace{1.cm}
   s~\tilde{F}_{0}^{L}(x) + U'(x)~\frac{\rmd\tilde{F}_{0}^{L}(x)}{\rmd x} -
   D~\frac{\rmd^{2}\tilde{F}_{0}^{L}(x)}{\rmd x^2} &=& 1, \\
   \label{eq20}
   \fl \hspace{1.cm}
   s ~\tilde{F}_{n}^{L}(x) + U'(x)~\frac{\rmd \tilde{F}_{n}^{L}(x)}{\rmd x}
   - D~\frac{\rmd^{2}\tilde{F}_{n}^{L}(x)}{\rmd x^{2}} &=&
   \frac{1}{\tau_{\rmd}}~\rme^{-(t'-t_{0})/\tau_{\rmd}}~\frac{\rmd}{\rmd x}\left[
   \tilde{F}_{n-1}^{L}(x)\Big\rfloor_{s-1/\tau_{\rmd}}\right].
\end{eqnarray}

\indent The Laplace transform of each term of the survival
probability, $\mathcal{L} [F_{n}(x,\tau;t')]$, is represented by
$\tilde{F}_{n}^{L}(x)$, omitting the parametric dependence on $s$
and tentatively on $t'$. In Eqs.~(\ref{eq19}) and~(\ref{eq20}),
the initial condition required by the Laplace transform of the
temporal derivative, $\mathcal{L}[\partial F_{n}(x,\tau;t') /
\partial \tau] = s ~\tilde{F}_{n}^{L}(x) - F_{n}(x,\tau=0;t')$,
has been assigned to an internal point, $x<x_{\rm thr}$, according
to Eqs.~(\ref{eq17}) and~(\ref{eq18}). This system of recursive
equations has to be solved together with the Laplace-transformed
boundary condition, $\tilde{F}_{n}^{L}(x=x_{\rm thr}) = 0$ for
all $n$.\\
\indent Up to now, we have addressed the evolution of survival
probability from the backward state, $x$ at time $t$, to the
current position, $x'$ at time $t'$. Since the experimental setup
-and in particular the temporally inhomogeneous exponential
time-dependent drift- refers to the time $t_0$, when the particle
is at position $x_0$, we need to link this survival probability to
the real initial state. According to the inverse Laplace
transform, the zeroth-order function of this probability

\begin{eqnarray}\label{eq21}
   F_{0}(x,\tau)= \frac{1}{2\pi {\rm j}}
   \int_{\sigma-{\rm j}\infty}^{\sigma+{\rm j}\infty}\rme^{s\tau}~
   \tilde{F}_{0}^{L}(x;s) ~\rmd s,
\end{eqnarray}

\noindent can be directly evaluated at the initial setting, $x =
x_0$ at time difference $\tau = t'-t_0$. In virtue of
Eq.~(\ref{eq19}), its Laplace transform satisfies

\begin{equation}
   \label{eq22}
   \frac{\rmd^{2}\tilde{F}_{0}^{L}(x_0)}{\rmd x_0^2} -
   \frac{U'(x_0)}{D}~\frac{\rmd\tilde{F}_{0}^{L}(x_0)}{\rmd x_0} -
   \frac{s}{D}~\tilde{F}_{0}^{L}(x_0) = -\frac{1}{D}
\end{equation}

\noindent and $\tilde{F}_{0}^{L}(x_0 = x_{\rm thr}) = 0$.\\
\indent From Eq.~(\ref{eq20}), it is easy to show that higher
order functions can be written as

\begin{equation}
   \label{eq23}
   \tilde{F}_{n}^{L}(x) = \rme^{-n(t'-t_{0})/\tau_{\rmd}}~
   \tilde{\mathbb{F}}_{n}^{L}(x),
\end{equation}

\noindent where the time-inhomogeneous part of the solution is
restricted to the exponential factor. The time-homogeneous
function $\tilde{\mathbb{F}}_{n}^{L}(x)$ satisfies

\begin{equation}
   \label{eq24}
   \frac{\rmd^{2}\tilde{\mathbb{F}}_{n}^{L}(x)}{\rmd x^2} -
   \frac{U'(x)}{D}~\frac{\rmd\tilde{\mathbb{F}}_{n}^{L}(x)}{\rmd x} -
   \frac{s}{D}~\tilde{\mathbb{F}}_{n}^{L}(x) = -\frac{1}{\tau_{\rmd}D}
   \frac{\rmd}{\rmd x}\tilde{\mathbb{F}}_{n-1}^{L}(x)\rfloor_{s-1/\tau_{\rmd}}.
\end{equation}

\indent To impose the real initial condition we need to obtain
each term in the temporal domain. According to the inverse Laplace
transform and Eq.~(\ref{eq23}), these functions read

\begin{eqnarray}\label{eq25}
   F_{n}(x,\tau;t')= \rme^{-n(t'-t_{0})/\tau_{\rmd}}~\frac{1}{2\pi {\rm j}}
   \int_{\sigma-{\rm j}\infty}^{\sigma+{\rm j}\infty}\rme^{s\tau}~
   \tilde{\mathbb{F}}_{n}^{L}(x;s) ~\rmd s.
\end{eqnarray}

\indent However, this integral cannot be done directly and certain
considerations have to be taken into account. Since we focus on
systems in which there are appropriate solutions to the FPT
problem in the unperturbed situation, the region of convergence of
$\tilde{F}_{0}^{L}(x)$, bounded by the path of integration in
Eq.~(\ref{eq21}), is assumed to be defined by $\sigma
> 0$. Consequently, because of the shift in $s$ in the forcing
term, the equation for the first-order function
$\tilde{\mathbb{F}}_{1}^{L}(x)$, Eq.~(\ref{eq24}) for $n=1$, is
valid in the region $\sigma > 1/\tau_{\rmd}$. The argument is
recursive and, therefore, the region of convergence of the
$n$th-order function is $\sigma > n/\tau_{\rmd}$. Accordingly, to
evaluate the integral in Eq.~(\ref{eq25}) we make the substitution
$z = s-n/\tau_{\rmd}$,

\begin{eqnarray}\label{eq26}
   F_{n}(x,\tau;t')= \rme^{-n(t'-t_{0})/\tau_{\rmd}}~\rme^{n\tau/\tau_{\rmd}}~
   \frac{1}{2\pi {\rm j}}\int_{\sigma_{z}-{\rm j}\infty}^{\sigma_{z}+{\rm j}\infty}
   \rme^{z\tau}~\tilde{\mathbb{F}}_{n}^{L}(x;z+n/\tau_{\rmd})
   ~\rmd z,
\end{eqnarray}

\noindent where now $\sigma_{z} > 0$. Once we evaluate the initial
setting, $x_0$ at time $t = t_0$ (or, equivalently, $\tau =
t'-t_0$), the exponential time-dependent factors cancel out and
the remaining contribution to the survival probability does not
depend on $t'$,

\begin{eqnarray}\label{eq27}
   F_{n}(x_0,\tau)= \frac{1}{2\pi {\rm j}}\int_{\sigma_{z}-{\rm j}\infty}
   ^{\sigma_{z}+{\rm j}\infty} \rme^{z\tau}~\tilde{\mathbb{F}}_{n}^{L}
   (x_0;z+n/\tau_{\rmd}) ~\rmd z.
\end{eqnarray}

\indent Moreover, the region of convergence of its Laplace
transform is the semi-plane defined by ${\rm Re}(s)
> 0$. To obtain $\tilde{\mathbb{F}}_{n}^{L} (x_0;z+n/\tau_{\rmd})$
we take the inverse Laplace transform on both sides of
Eq.~(\ref{eq24}),

\begin{eqnarray}
   \label{eq28}
   \fl \hspace{1.cm}
   \frac{1}{2\pi {\rm j}} \int_{\sigma-{\rm j}\infty}^{\sigma+{\rm j}\infty}
   \rme^{s\tau}~ \left[ \frac{\rmd^{2}\tilde{\mathbb{F}}_{n}^{L}(x_0;s)}{\rmd x_0^2} -
   \frac{U'(x_0)}{D}~\frac{\rmd\tilde{\mathbb{F}}_{n}^{L}(x_0;s)}{\rmd x_0} -
   \frac{s}{D}~\tilde{\mathbb{F}}_{n}^{L}(x_0;s) \right] ~ \rmd s \nonumber\\
   = -\frac{1}{\tau_{\rmd}D} ~\frac{1}{2\pi {\rm j}} \int_{\sigma-{\rm j}\infty}
   ^{\sigma+{\rm j}\infty} \rme^{s\tau}~\frac{\rmd}{\rmd x_0}\tilde{\mathbb{F}}
   _{n-1}^{L}(x_0;s-1/\tau_{\rmd}) ~ \rmd s,
\end{eqnarray}

\noindent and apply the same substitution as before, $z =
s-n/\tau_{\rmd}$,

\begin{eqnarray}
   \label{eq29}
   \fl \hspace{1.cm}
   \frac{1}{2\pi {\rm j}} \int_{\sigma_z-{\rm j}\infty}^{\sigma_z+{\rm j}\infty}
   \rme^{z\tau}~ \left[ \frac{\rmd^{2}\tilde{\mathbb{F}}_{n}^{L}}{\rmd x_0^2}
   - \frac{U'(x_0)}{D}~\frac{\rmd\tilde{\mathbb{F}}_{n}^{L}}{\rmd x_0}
   - \frac{(z+n/\tau_{\rmd})}{D}~\tilde{\mathbb{F}}_{n}^{L} \right] ~ \rmd z
   \nonumber\\
   = -\frac{1}{\tau_{\rmd}D} ~\frac{1}{2\pi {\rm j}}
   \int_{\sigma_z-{\rm j}\infty} ^{\sigma_z+{\rm j}\infty} \rme^{z\tau}~
   \frac{\rmd}{\rmd x_0}\tilde{\mathbb{F}}_{n-1}^{L} ~ \rmd z,
\end{eqnarray}

\noindent where we have simplified the notation to
$\tilde{\mathbb{F}}_{n}^{L} = \tilde{\mathbb{F}}_{n}^{L}
(x_0;z+n/\tau_{\rmd})$. From Eq.~(\ref{eq29}), it is
straightforward to show that each function $\tilde{\mathbb{F}}
_{n}^{L}(x_0;z+n/\tau_{\rmd})$ appearing in the integrand of
Eq.~(\ref{eq27}) satisfies

\begin{equation}
   \label{eq30}
   \fl \hspace{1.0cm}
   \frac{\rmd^{2}\tilde{\mathbb{F}}_{n}^{L}(x_0)}{\rmd x_0^2}
   - \frac{U'(x_0)}{D}~\frac{\rmd\tilde{\mathbb{F}}_{n}^{L}(x_0)}{\rmd x_0}
   - \frac{(s+n/\tau_{\rmd})}{D}~\tilde{\mathbb{F}}_{n}^{L}(x_0)
   = -\frac{1}{\tau_{\rmd}D} ~ \frac{\rmd}{\rmd x_0}
   \tilde{\mathbb{F}}_{n-1}^{L}(x_0),
\end{equation}

\noindent with $\tilde{\mathbb{F}}_{n}^{L}(x_0 = x_{\rm thr}) =
0$.\\
\indent To summarize, the survival probability from the initial
state $(x_0,t_0)$ to the current state $(x,t_0 + \tau)$ is given
by

\begin{equation}\label{eq31}
   \fl \hspace{0.5cm}
   F(x_0,\tau) = F_{0}(x_0,\tau) + \epsilon~F_{1}(x_0,\tau) +
   \epsilon^{2}~F_{2}(x_0,\tau) + \dots =
   \sum_{n=0}^{\infty}\epsilon^{n}~F_{n}(x_0,\tau),
\end{equation}

\noindent where all functions are obtained from the corresponding
inverse Laplace transforms, either given by Eq.~(\ref{eq21}) or
Eq.~(\ref{eq27}). In turn, the Laplace transform of the
unperturbed system, $\tilde{F}_{0}^{L}(x_0)$, satisfies
Eq.~(\ref{eq22}), whereas all time-homogeneous higher order terms,
$\tilde{\mathbb{F}}_{n}^{L}(x_0)$, are recursively obtained from
Eq.~(\ref{eq30}). For any order, the boundary condition is zero,
$\tilde{F}_{0}^{L}(x_{\rm thr}) = \tilde{\mathbb{F}}_{n}
^{L}(x_{\rm thr}) = 0$.\\
\indent As shown in Section \ref{unperturbed}, FPT density
function can be directly obtained once the survival probability
from the initial state to the current state is known, see
Eq.~(\ref{eq9}). Therefore, the series structure in
Eq.~(\ref{eq31}) is inherited by the FPT density function
\cite{Urdapilleta2011a, Urdapilleta2012},

\begin{equation}\label{eq32}
   \phi(x_0,\tau) = \sum_{n=0}^{\infty}\epsilon^{n}~\phi_{n}(x_0,\tau),
\end{equation}

\noindent where each order function, $\phi_{n}(x_0,\tau)$,
satisfies

\begin{equation}\label{eq33}
   \phi_{n}(x_0,\tau) = - \frac{\partial F_{n}(x_0,\tau)}{\partial \tau}.
\end{equation}

\indent In terms of the Laplace transform, this set of equations
reads

\begin{eqnarray}\label{eq34}
   \tilde{\phi}_{0}^{L}(x_0;s) = 1 - s ~ \tilde{F}_{0}^{L}(x_0;s) \\
   \label{eq35}
   \tilde{\phi}_{n}^{L}(x_0;s) = - s ~ \tilde{\mathbb{F}}_{n}^{L}(x_0;s),
   ~~~~n \geq 1,
\end{eqnarray}

\noindent where we have taken into account the conditions given by
Eqs.~(\ref{eq17}) and (\ref{eq18}) at the real initial state. With
these relationships, and based on Eqs.~(\ref{eq22}) and
(\ref{eq30}), it is straightforward to derive the system of
equations governing each term in the series solution for the FPT
density function, Eq.~(\ref{eq32}),

\begin{eqnarray}\label{eq36}
   \fl \hspace{1.75cm}
   \frac{\rmd^{2}\tilde{\phi}_{0}^{L}(x_0)}{\rmd x_0^2} -
   \frac{U'(x_0)}{D}~\frac{\rmd\tilde{\phi}_{0}^{L}(x_0)}{\rmd x_0} -
   \frac{s}{D}~\tilde{\phi}_{0}^{L}(x_0) &=& 0, \\
   \label{eq37}
   \fl \hspace{0.25cm}
   \frac{\rmd^{2}\tilde{\phi}_{n}^{L}(x_0)}{\rmd x_0^2}
   - \frac{U'(x_0)}{D}~\frac{\rmd\tilde{\phi}_{n}^{L}(x_0)}{\rmd x_0}
   - \frac{(s+n/\tau_{\rmd})}{D}~\tilde{\phi}_{n}^{L}(x_0)
   &=& -\frac{1}{\tau_{\rmd}D} ~ \frac{\rmd}{\rmd x_0}
   \tilde{\phi}_{n-1}^{L}(x_0), ~n \geq 1,
\end{eqnarray}

\noindent with boundary conditions $\tilde{\phi}_{0}^{L}(x_{\rm
thr}) = 1$ and $\tilde{\phi}_{n}^{L}(x_{\rm thr}) = 0$ for $n\geq
1$. In all cases, a second boundary condition is necessary for
these second order ordinary differential equations; as usual,
bounded solutions at $x_0 \rightarrow -\infty$ are required.

\section{Explicit solutions}
\subsection{FPT statistics for the Wiener process}\label{wiener}
\indent The Wiener process is defined by the potential $U(x) =
-\mu~x$, where $\mu > 0$ guarantees threshold crossing. The
associated Langevin equation with additive time-dependent
exponential drift is

\begin{equation}\label{eq38}
   \frac{\rmd x}{\rmd t} = \mu + \frac{\epsilon}
   {\tau_{\rmd}}~\rme^{-(t-t_0)/\tau_{\rmd}} + \sqrt{2D}~\xi(t).
\end{equation}

\indent This is the simplest system for the kinds of diffusion
processes we are considering and it has already been solved in
\cite{Urdapilleta2011a, Urdapilleta2012}. The formulation of the
FPT problem in these works is equivalent to the derivation
presented here, although in this article we have managed to extend
it to a general context, without explicit knowledge of the
survival probability.\\
\indent The solution to Eqs.~(\ref{eq36}) and (\ref{eq37}) for
this particular potential read

\begin{eqnarray}
   \label{eq39}
   \fl \hspace{0.5cm}
   \tilde{\phi}_{0}^{L}(x_0;s) = \exp\left\{\frac{(x_{\rm thr}-x_{0})}
   {2D}[\mu-\sqrt{\mu^{2}+4Ds}]\right\},\\
   \label{eq40}
   \fl \hspace{0.5cm}
   \tilde{\phi}_{n}^{L}(x_0;s) = - \frac{[\mu-\sqrt{\mu^{2}+4Ds}]}{2D}\nonumber\\
   \fl \hspace{1.75cm}
   \times~\sum_{k=0}^{n}b_{n,k}(s)~\exp\left\{
   \frac{(x_{\rm thr}-x_{0})}{2D}[\mu-\sqrt{\mu^{2}+4D(s+k/\tau_{\rm
   d})}]\right\},~~{\rm for}~n\geq 1,
\end{eqnarray}

\noindent where the coefficients are given by the recursive
scheme,

\begin{eqnarray}\label{eq41}
   \fl \hspace{0.5cm}
   b_{n,k}(s) &=&
   -\frac{b_{n-1,k}(s)}{n-k}~\frac{[\mu-\sqrt{\mu^{2}+4D(s+k/\tau_{\rm d})}]}{2D},~~{\rm for}~k=0,\dots,n-1,\\
   \fl \hspace{0.5cm}
   \label{eq42}
   b_{n,n}(s) &=& -\sum_{k=0}^{n-1} b_{n,k}(s),
\end{eqnarray}

\noindent starting from $b_{1,0}(s)=1$ and $b_{1,1}(s)=-1$. As
shown in \cite{Urdapilleta2012}, it can be shown by mathematical
induction that these expressions satisfy Eqs.~(\ref{eq36}) and
(\ref{eq37}), for $U'(x_0) = -\mu$.

\subsection{FPT statistics for the Ornstein-Uhlenbeck process}\label{ou}
\indent The potential for the Ornstein-Uhlenbeck process is given
by $U(x) = -\mu~x + \frac{1}{2} ~ x^2 / \tau_{\rm m}$, where the
time constant $\tau_{\rm m}$ characterizes the relaxation process
towards the equilibrium point, $x_{\rm eq} = \mu ~ \tau_{\rm m}$,
for noise-free dynamics without a threshold. Alternatively,
$\tau_{\rm m}$ sets the time constant of the exponential
autocorrelation function of this process. The presence of the
threshold divides the behavior of the system according to the
driving force $\mu$. In the \textit{supra-threshold} regime (also
denoted \textit{stimulus-driven}), $\mu > x_{\rm thr} / \tau_{\rm
m}$, whenever the particle is released at $x_{0} < x_{\rm thr}$,
it reaches the threshold in a finite time. In contrast, in the
\textit{sub-threshold} regime (or \textit{noise-driven}), $\mu <
x_{\rm thr} / \tau_{\rm m}$, the particle cannot reach the
threshold unless assisted by noise. In the special case $\mu =
x_{\rm thr} / \tau_{\rm m}$, the particle hits the threshold in an
infinite time. When supplemented by the time-dependent exponential
drift, the associated Langevin equation reads

\begin{equation}\label{eq43}
   \frac{\rmd x}{\rmd t} = \mu - \frac{x}{\tau_{\rm m}} + \frac{\epsilon}
   {\tau_{\rmd}}~\rme^{-(t-t_0)/\tau_{\rmd}} + \sqrt{2D}~\xi(t).
\end{equation}

\subsubsection{Zeroth-order density function}
\indent The density function of the FPT for the unperturbed
system, $\phi_{0}(x_0;\tau)$, can be obtained from the inverse
Laplace transform of the solution to Eq.~(\ref{eq36}) for this
particular potential; in detail,

\begin{equation}\label{eq44}
   \frac{\rmd^{2}\tilde{\phi}_{0}^{L}}{\rmd x_0^2} +
   \left( \frac{\mu}{D} - \frac{x_0}{\tau_{\rm m} D} \right) ~
   \frac{\rmd\tilde{\phi}_{0}^{L}}{\rmd x_0} -
   \frac{s}{D}~\tilde{\phi}_{0}^{L} = 0.
\end{equation}

\indent It can be shown that, with the change of variable

\begin{equation}\label{eq45}
   z = \sqrt{\frac{\tau_{\rm m}}{D}}\left( \mu - \frac{x_0}{\tau_{\rm m}}
    \right),
\end{equation}

\noindent the solution can be expressed as \cite{Darling1953,
Roy1969, Capocelli1971}

\begin{equation}\label{eq46}
   \tilde{\phi}_{0}^{L}(x_0;s) = \rme^{z^2 / 4}~u_0(z;s),
\end{equation}

\noindent where $u_0(z;s)$ satisfies

\begin{equation}\label{eq47}
   \frac{\rmd^{2}u_0}{\rmd z^2} +
   \left( -\tau_{\rm m}~s + \frac{1}{2} - \frac{1}{4} z^2 \right)
   ~u_0 = 0.
\end{equation}

\indent The general solution to this homogeneous equation reads

\begin{equation}\label{eq48}
   u_0(z;s) = c_1 ~ \mathcal{D}_{-\tau_{\rm m}s}(z) + c_2 ~
   \mathcal{D}_{-\tau_{\rm m}s}(-z),
\end{equation}

\noindent where $\mathcal{D}_\nu(z)$ are the parabolic cylinder
functions according to Whittaker's notation \cite{Handbook}. Given
that $\mathcal{D}_{-\tau_{\rm m}s}(-z)$ diverges at $z \rightarrow
\infty$, well-behaved solutions require that $c_2 = 0$. Note that
because of the transformation, the new domain is from $z_{\rm
thr}$ to $\infty$, where $z_{\rm thr} = \sqrt{\tau_{\rm m}/D}~(
\mu - x_{\rm thr}/\tau_{\rm m})$. In virtue of the boundary
condition at the threshold, $\tilde{\phi}_{0}^{L} (x_{\rm thr}) =
1$, it is straightforward to show that

\begin{equation}\label{eq49}
   u_0(z;s) = \frac{\rme^{-z_{\rm thr}^2 / 4}}{\mathcal{D}_
   {-\tau_{\rm m}s}(z_{\rm thr})}~\mathcal{D}_{-\tau_{\rm m}s}(z).
\end{equation}

\indent Equivalently,

\begin{equation}\label{eq50}
   \tilde{\phi}_{0}^{L}(x_0;s) = {\rme}^{-\frac{\tau_{\rm
   m}}{4D}\left[ \left(\mu-\frac{x_{\rm thr}}{\tau_{\rm m}}\right)^2 -
   \left(\mu-\frac{x_0}{\tau_{\rm m}}\right)^2 \right]}~\frac{\mathcal{D}_
   {-\tau_{\rm m}s}\left[ \sqrt{\frac{\tau_{\rm m}}{D}}\left( \mu -
   \frac{x_0}{\tau_{\rm m}} \right) \right]}{\mathcal{D}_
   {-\tau_{\rm m}s}\left[ \sqrt{\frac{\tau_{\rm m}}{D}}\left( \mu -
   \frac{x_{\rm thr}}{\tau_{\rm m}} \right) \right]}.
\end{equation}

\subsubsection{Higher-order density functions}
\indent According to Eq.~(\ref{eq37}), the Laplace transform of
higher-order density functions, $\tilde{\phi}_{n}^{L}(x_0;s)$, for
the Ornstein-Uhlenbeck process satisfies

\begin{equation}\label{eq51}
   \fl \hspace{1.cm}
   \frac{\rmd^{2}\tilde{\phi}_{n}^{L}}{\rmd x_0^2}
   + \left( \frac{\mu}{D} - \frac{x_0}{\tau_{\rm m} D} \right) ~
   \frac{\rmd\tilde{\phi}_{n}^{L}}{\rmd x_0}
   - \frac{(s+n/\tau_{\rmd})}{D}~\tilde{\phi}_{n}^{L}
   = -\frac{1}{\tau_{\rmd}D} ~ \frac{\rmd \tilde{\phi}_{n-1}^{L}}
   {\rmd x_0},  ~n \geq 1.
\end{equation}

\indent In terms of the variable $z$, defined by Eq.~(\ref{eq45}),
it is easy to show that all functions can be written as

\begin{equation}\label{eq52}
   \tilde{\phi}_{n}^{L}(x_0;s) = \rme^{z^2 / 4}~u_{n}(z;s),
\end{equation}

\noindent where $u_{n}(z;s)$, for $n \geq 1$, is governed by

\begin{equation}\label{eq53}
   \fl \hspace{1.cm}
   \frac{\rmd^{2}u_{n}}{\rmd z^2} +
   \left[ -\tau_{\rm m}\left( s + \frac{n}{\tau_{\rmd}} \right)
   + \frac{1}{2} - \frac{1}{4} z^2 \right]~u_{n} = \frac{1}
   {\tau_{\rmd}}~\sqrt{\frac{\tau_{\rm m}}{D}}~\left(
   \frac{1}{2}~z~u_{n-1} + \frac{\rmd u_{n-1}}{\rmd z} \right),
\end{equation}

\noindent and the boundary condition at the transformed threshold
is $u_{n}(z_{\rm thr};s) = 0$.\\
\indent Next, we prove by mathematical induction that for
$\tau_{\rm m} \neq \tau_{\rmd}$, the solution to Eq.~(\ref{eq53})
reads

\begin{equation}\label{eq54}
   u_{n}(z;s) = \sum_{k=0}^{n} b_{n,k}(s)~\mathcal{D}_{-\tau_{\rm m}
   [s+(n-k)/\tau_{\rm m}+k/\tau_{\rm d}]}(z),
\end{equation}

\noindent where the coefficients are given by the recursive
structure,

\begin{eqnarray}
   \label{eq55}
   \fl
   b_{n,k}(s) &=& \frac{[s+(n-1-k)/\tau_{\rm m} + k/\tau_{\rmd}]}
   {(n-k)}~\frac{\sqrt{\tau_{\rm m}/D}}{(1-\tau_{\rmd}/\tau_{\rm m})}
   ~b_{n-1,k}(s),~~{\rm for}~k=0, \dots, n-1,\\
   \label{eq56}
   \fl
   b_{n,n}(s) &=& - \frac{1}{\mathcal{D}_{-\tau_{\rm m}(s+n/\tau_{\rmd})}
   (z_{\rm thr})} \sum_{k=0}^{n-1} b_{n,k}(s)~\mathcal{D}_
   {-\tau_{\rm m}[s+(n-k)/\tau_{\rm m}+k/\tau_{\rmd}]}(z_{\rm thr}),
\end{eqnarray}

\noindent starting from

\begin{eqnarray}
   \label{eq57}
   b_{1,0}(s) &=& \frac{\sqrt{\tau_{\rm m}/D}}{(1-\tau_{\rmd}/\tau_{\rm m})}~
   \frac{\rme^{-z_{\rm thr}^2 / 4}}{\mathcal{D}_{-\tau_{\rm m}s}
   (z_{\rm thr})}~s,\\
   \label{eq58}
   b_{1,1}(s) &=& - \frac{\sqrt{\tau_{\rm m}/D}}{(1-\tau_{\rmd}/\tau_{\rm m})}~
   \frac{\rme^{-z_{\rm thr}^2 / 4}}{\mathcal{D}_{-\tau_{\rm m}s}
   (z_{\rm thr})}~\frac{\mathcal{D}_{-\tau_{\rm m}(s+1/\tau_{\rm m})}(z_{\rm thr})}
   {\mathcal{D}_{-\tau_{\rm m}(s+1/\tau_{\rmd})}(z_{\rm thr})}~s.
\end{eqnarray}

\indent Assuming that the $n$th-order function is given by
Eq.~(\ref{eq54}), according to Eq.~(\ref{eq53}) the
$(n+1)$th-order function should satisfy

\begin{eqnarray}
   \fl \hspace{0.5cm}
   \frac{\rmd^{2}u_{n+1}}{\rmd z^2} +
   \left\{ -\tau_{\rm m}\left[ s + \frac{(n+1)}{\tau_{\rmd}} \right]
   + \frac{1}{2} - \frac{1}{4} z^2 \right\}~u_{n+1} = \frac{1}{\tau_{\rmd}} ~
   \sqrt{\frac{\tau_{\rm m}}{D}}\nonumber\\
   \label{eq59}
   \fl \hspace{1.5cm}
   \times \sum_{k=0}^{n} b_{n,k}(s)~
   \left\{ \frac{1}{2}~z~\mathcal{D}_{-\tau_{\rm m}
   [s+(n-k)/\tau_{\rm m}+k/\tau_{\rm d}]}(z) + \frac{\rmd
   \mathcal{D}_{-\tau_{\rm m}[s+(n-k)/
   \tau_{\rm m}+k/\tau_{\rm d}]}(z)}{\rmd z} \right\}.
\end{eqnarray}

\indent From the recurrence relationships that Weber parabolic
cylinder functions satisfy \cite{Handbook}, it is easy to show
that

\begin{equation}\label{eq60}
   \frac{\rmd \mathcal{D}_{\nu}(z)}{\rmd z} + \frac{1}{2}~z~
   \mathcal{D}_{\nu}(z) = \nu~\mathcal{D}_{\nu-1}(z),
\end{equation}

\noindent and, therefore, the forcing term in Eq.~(\ref{eq59})
simplifies to

\begin{eqnarray}
   \fl \hspace{0.5cm}
   \frac{\rmd^{2}u_{n+1}}{\rmd z^2} +
   \left\{ -\tau_{\rm m}\left[ s + \frac{(n+1)}{\tau_{\rmd}} \right]
   + \frac{1}{2} - \frac{1}{4} z^2 \right\}~u_{n+1} = -\frac{\tau_{\rm m}}
   {\tau_{\rmd}} ~ \sqrt{\frac{\tau_{\rm m}}{D}} \nonumber\\
   \label{eq61}
   \fl \hspace{1.5cm}
   \times \sum_{k=0}^{n} b_{n,k}(s)~\left[ s + \frac{(n-k)}
   {\tau_{\rm m}} + \frac{k}{\tau_{\rm d}} \right]~
   \mathcal{D}_{-\tau_{\rm m}[s+(n+1-k)/\tau_{\rm m}+
   k/\tau_{\rm d}]}(z).
\end{eqnarray}

\indent The solution to the homogeneous part of this equation is
given by an analogous expression to Eq.~(\ref{eq48}),

\begin{equation}\label{eq62}
   \fl \hspace{0.75cm}
   u_{n+1}^{\rm hom.}(z;s) = c_1 ~ \mathcal{D}_{-\tau_{\rm m}[s+(n+1)/\tau_{\rm d}]}(z) + c_2 ~
   \mathcal{D}_{-\tau_{\rm m}[s+(n+1)/\tau_{\rm d}]}(-z),
\end{equation}

\noindent whereas a particular solution is

\begin{eqnarray}
   \fl \hspace{0.75cm}
   u_{n+1}^{\rm part.}(z;s) = \frac{\sqrt{\tau_{\rm m}/D}}{(1-\tau_{\rmd}/\tau_{\rm m})}
   \sum_{k=0}^{n} \frac{[s+(n-k)/\tau_{\rm m}+k/\tau_{\rm d}]}{(n+1-k)}
   ~b_{n,k}(s)~ \nonumber\\
   \label{eq63}
   \fl \hspace{8.5cm}
   \times \mathcal{D}_{-\tau_{\rm m}[s+(n+1-k)/\tau_{\rm m}+
   k/\tau_{\rm d}]}(z).
\end{eqnarray}

\indent The recursive structure of the coefficients is implicitly
defined in Eq.~(\ref{eq63}) and agrees to Eq.~(\ref{eq55}).
Therefore, the general solution $u_{n+1}(z;s) = u_{n+1}^{\rm
hom.}(z;s) + u_{n+1}^{\rm part.}(z;s)$ can be expressed as

\begin{eqnarray}
   \fl \hspace{0.75cm}
   u_{n+1}(z;s) = \sum_{k=0}^{n} b_{n+1,k}(s)~ \mathcal{D}_{-\tau_{\rm m}
   [s+(n+1-k)/\tau_{\rm m}+k/\tau_{\rm d}]}(z) + c_1 ~ \mathcal{D}_{-\tau_{\rm m}
   [s+(n+1)/\tau_{\rm d}]}(z) \nonumber\\
   \label{eq64}
   \fl \hspace{7.cm}
   + ~c_2 ~ \mathcal{D}_{-\tau_{\rm m} [s+(n+1)/\tau_{\rm d}]}(-z).
\end{eqnarray}

\indent As before, an appropriate behavior at $z \rightarrow
-\infty$ implies that $c_2 = 0$, and the evaluation of the
boundary condition, $u_{n+1}(z_{\rm thr};s) = 0$, determines
$c_1$. Taking into account the definition given in
Eq.~(\ref{eq56}), the $(n+1)$th-order function reads

\begin{eqnarray}
   \label{eq65}
   \fl \hspace{0.75cm}
   u_{n+1}(z;s) = \sum_{k=0}^{n} b_{n+1,k}(s)~ \mathcal{D}_{-\tau_{\rm m}
   [s+(n+1-k)/\tau_{\rm m}+k/\tau_{\rm d}]}(z) \nonumber\\
   \fl \hspace{7.cm}
   + ~b_{n+1,n+1}(s)~\mathcal{D}_{-\tau_{\rm m}[s+(n+1)/\tau_{\rm
   d}]}(z), \nonumber\\
   \fl \hspace{2.6cm}
   = \sum_{k=0}^{n+1} b_{n+1,k}(s)~ \mathcal{D}_{-\tau_{\rm m}
   [s+(n+1-k)/\tau_{\rm m}+k/\tau_{\rm d}]}(z),
\end{eqnarray}

\noindent which clearly satisfies Eq.~(\ref{eq54}) for $(n+1)$.
Therefore, as far as Eq.~(\ref{eq54}) is true for the $n$th-order
function, it holds for the following order, where coefficients are
related by Eqs.~(\ref{eq55}) and (\ref{eq56}).\\
\indent The proof is completed by observing that the first-order
function $u_{1}(z;s)$ belongs to the family described by
Eq.~(\ref{eq54}). According to Eq.~(\ref{eq53}) and taking into
account the solution for the zeroth-order function,
Eq.~(\ref{eq49}), this function satisfies

\begin{equation}\label{eq66}
   \fl
   \frac{\rmd^{2}u_{1}}{\rmd z^2} +
   \left[ -\tau_{\rm m} \left(s + \frac{1}{\tau_{\rmd}} \right)
   + \frac{1}{2} - \frac{1}{4} z^2 \right]~u_{1} = -\frac{\tau_{\rm m}}
   {\tau_{\rmd}} ~ \sqrt{\frac{\tau_{\rm m}}{D}}~\frac{\rme^{-z_{\rm thr}^2
   / 4}}{\mathcal{D}_{-\tau_{\rm m}s}(z_{\rm thr})}~s~\mathcal{D}_
   {-\tau_{\rm m}(s+1/\tau_{\rm m})}(z),
\end{equation}

\noindent where we have used the recurrence relationship given by
Eq.~(\ref{eq60}). The general solution to this equation is

\begin{eqnarray}
   \fl \hspace{0.75cm}
   u_1(z;s) = c_1~\mathcal{D}_{-\tau_{\rm m}(s+1/\tau_{\rm d})}(z)
   + c_2~\mathcal{D}_{-\tau_{\rm m}(s+1/\tau_{\rm d})}(-z) \nonumber\\
   \label{eq67}
   \fl \hspace{5.0cm}
   +\frac{\sqrt{\tau_{\rm m}/D}}{(1-\tau_{\rmd}/\tau_{\rm m})}~
   \frac{\rme^{-z_{\rm thr}^2 / 4}}{\mathcal{D}_{-\tau_{\rm m}s}
   (z_{\rm thr})}~s~\mathcal{D}_{-\tau_{\rm m}(s+1/\tau_{\rm
   m})}(z).
\end{eqnarray}

\indent Since $c_2 = 0$ for bounded solutions and $c_1$ is
determined from the evaluation of the boundary condition,
$u_{1}(z_{\rm thr};s) = 0$, the first-order function is explicitly
given by

\begin{eqnarray}
   \fl \hspace{0.75cm}
   u_1(z;s) = \frac{\sqrt{\tau_{\rm m}/D}}{(1-\tau_{\rmd}/\tau_{\rm m})}~
   \frac{\rme^{-z_{\rm thr}^2 / 4}}{\mathcal{D}_{-\tau_{\rm m}s}
   (z_{\rm thr})}~s~\mathcal{D}_{-\tau_{\rm m}(s+1/\tau_{\rm
   m})}(z) \nonumber\\
   \label{eq68}
   \fl \hspace{2.5cm}
   -  \frac{\sqrt{\tau_{\rm m}/D}}{(1-\tau_{\rmd}/\tau_{\rm m})}~
   \frac{\rme^{-z_{\rm thr}^2 / 4}}{\mathcal{D}_{-\tau_{\rm m}s}
   (z_{\rm thr})}~\frac{\mathcal{D}_{-\tau_{\rm m}(s+1/\tau_{\rm m})}(z_{\rm thr})}
   {\mathcal{D}_{-\tau_{\rm m}(s+1/\tau_{\rmd})}(z_{\rm thr})}~s~
   \mathcal{D}_{-\tau_{\rm m}(s+1/\tau_{\rm d})}(z),
\end{eqnarray}

\noindent which can be expressed according to Eq.~(\ref{eq54}),
with coefficients given by Eqs.~(\ref{eq57}) and (\ref{eq58}).
This step ends the demonstration by mathematical induction of the
series solution to the FPT density function, given by
Eqs.~(\ref{eq52}) and (\ref{eq54}) in the transformed variable $z$
with coefficients defined by Eqs.~(\ref{eq55})-(\ref{eq58}).\\
\indent For completeness, the recursive solution for the special
case of the Ornstein-Uhlenbeck process with $\tau_{\rm m} =
\tau_{\rmd}$ is given in the Appendix.

\subsection{Other potentials}
\indent The survival probability and FPT statistics for a system
driven by an exponential time-dependent drift in an arbitrary
potential can be \textit{formally} obtained in a recursive scheme,
whenever the Green's function of Eq.~(\ref{eq37}) exists. This
function is practically the same as that of the unperturbed
system, Eq.~(\ref{eq36}), so essentially the FPT problem for the
time-inhomogeneous setup is as analytically tractable as for the
unperturbed case. The procedure is exemplified in the Appendix,
for the case of the Ornstein-Uhlenbeck process with $\tau_{\rm m}
= \tau_{\rmd}$.

\section{Comparison to numerical simulations}
\indent As the solution we have found for the present FPT problem
is given in terms of a series, we need to confirm its convergence
or resort to numerical simulations to test its validity. We did
not come to any general conclusions about the domain of
convergence in the parameter space, and so here we have
exemplified the usefulness of our approach by taking several cases
spanning different regimes, for those systems for which we have
found explicit solutions. For the Wiener process, we refer the
reader to \cite{Urdapilleta2012}, where we have shown that the
series solution is valid in the prototypical case corresponding to
the supra-threshold regime, as it provides an excellent
description far beyond the linear perturbative scenario. For the
Ornstein-Uhlenbeck process, a number of parameters is available:
$\tau_{\rm m}$, $x_{\rm thr}$, $x_0$, $\mu$, $D$, $\tau_{\rm d}$,
and $\epsilon$. However, the exact values of $x_{\rm thr}$ and
$x_0$ do not add any complexity to the model, and $\tau_{\rm m}$
sets the timescale of the dynamics. Therefore, there are three
main parameters to explore, $\mu$, $D$, and $\tau_{\rm d}$,
whereas $\epsilon$ controls the intensity of the superimposed
time-dependent drift.

\begin{figure}[t!]
\begin{center}
\includegraphics[scale=1.0]{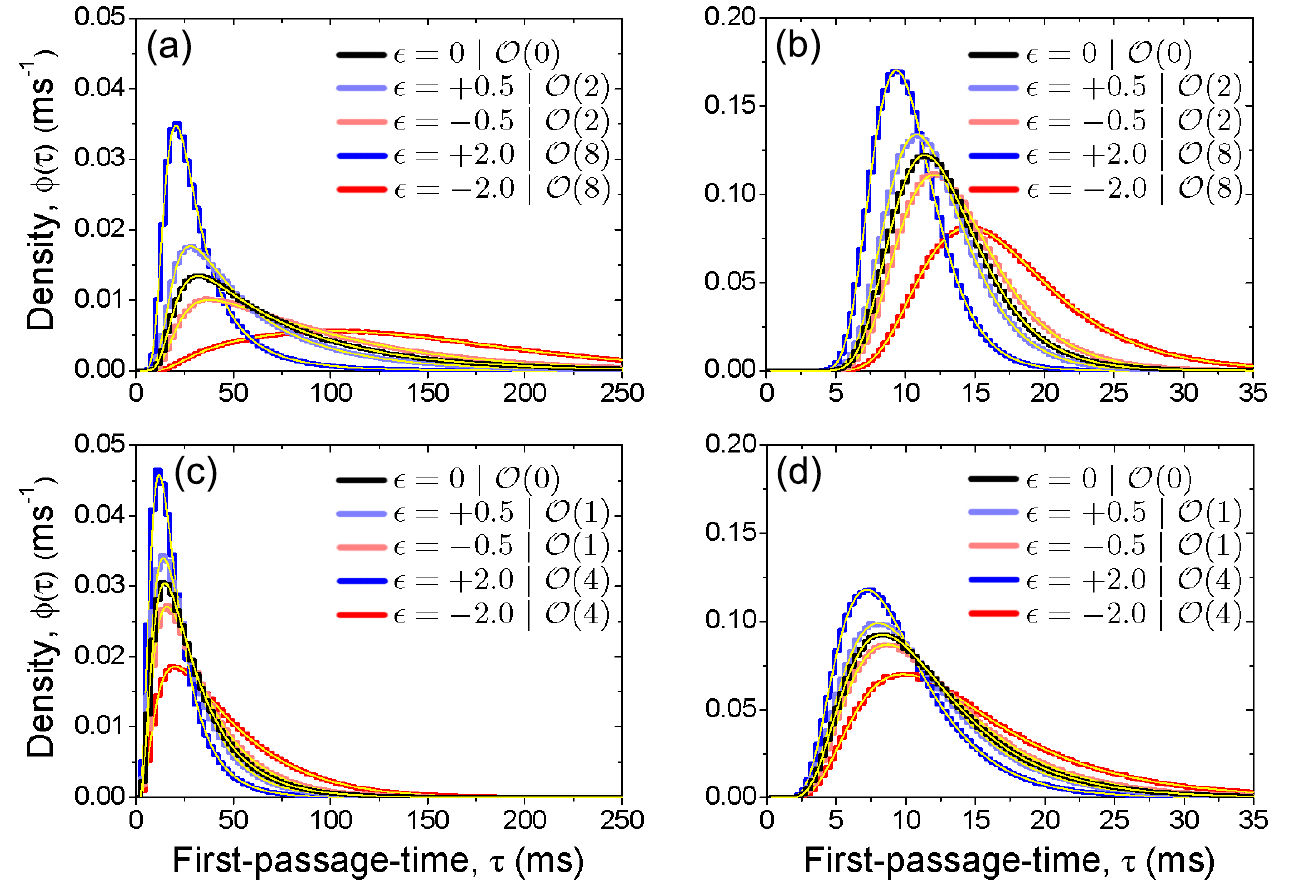}
\caption{\label{fig1} Comparison between the series solution for
the FPT density function and numerical results for different cases
of the Ornstein-Uhlenbeck process. (a) Low noise sub-threshold
regime. (b) Low noise supra-threshold regime. (c) High noise
sub-threshold regime. (d) High noise supra-threshold regime. In
all cases, $x_0 = 0$, $x_{\rm thr} = 1$, $\tau_{\rm m} = 10~{\rm
ms}$, and $\tau_{\rm d} = 100~{\rm ms}$. Different regimes are
defined by $\mu = 0.075~{\rm ms}^{-1}$ (sub-threshold), $\mu =
0.1333~{\rm ms}^{-1}$ (supra-threshold), $D = 0.0025~{\rm
ms}^{-1}$ (low noise), and $D = 0.01~{\rm ms}^{-1}$ (high noise).
FPT distributions obtained from the numerical simulation of
Eq.~(\ref{eq43}) for different intensities of the time-dependent
exponential drift are shown by thick stair-like colored lines,
labeled in the upper-right hand part of each panel: blue, light
blue, black, light red, and red correspond to $\epsilon = +2.0,
+0.5, 0, -0.5, -2.0$, respectively. The analytical expression,
Eq.~(\ref{eq32}), obtained from the numerical inverse Laplace
transform of the explicit solution, Eq.~(\ref{eq50}) for the
unperturbed density, and Eqs.~(\ref{eq52}) and (\ref{eq54}) for
higher order terms, up to the order $N$ indicated for each
intensity, $\mathcal{O}(N)$, is represented as a thin yellow line.
In all cases, the analytical results excellently describe FPT
statistics obtained from numerical simulations.}
\end{center}
\end{figure}

\indent Given our interest in neuronal adaptation (see Section
\ref{intro}), we focus on the correspondence between the
Ornstein-Uhlenbeck process and the \textit{leaky
integrate-and-fire} neuron model to define the typical parameters.
In this model, $x$ is the transmembrane voltage, the input derived
from the potential, $-U'(x)$, represents external driving as well
as currents flowing through specific (leaky) membrane channels,
and the superimposed exponential temporal drift corresponds to an
adaptation current. Without stimulation, voltage decays to the
resting potential, $x_0$, whereas when inputs drive the neuron
across the threshold, $x_{\rm thr}$, a spike is declared and the
voltage is reset to a starting point, here assumed to be equal to
$x_0$. The specific problem of neuronal adaptation also includes a
history-dependent process (see Section \ref{discussion} for
further discussion), not included here. By redefining
$(x-x_0)/x_{\rm thr} \rightarrow x$, the new dimensionless voltage
$x$ starts at $x_0 = 0$ and the FTP problem corresponds to reach
the threshold $x_{\rm thr} = 1$. On the other hand, the temporal
scale is set by $\tau_{\rm m}$, here assumed to be $\tau_{\rm m} =
10$ time units (ms), in agreement with the experimental values
\cite{Koch, GerstnerKistler}. All parameter values explored here
have to be compared to this particular $\tau_{\rm m}$; otherwise,
an adimensionalization procedure can be used. The remaining
parameters are used to explore the convergence of the series
solution in different cases. As mentioned in Section \ref{ou},
different dynamical regimes can be defined according to the
intensity of the constant drift, $\mu$. In turn, the intensity of
the noise, $D$, essentially modulates the dispersion of the FPT
distribution. Once the set of parameters to examine has been
defined, numerical FPT distributions are obtained from
first-hitting times of the system evolving according to the
Langevin equation that governs its dynamics, Eq.~(\ref{eq43}).

\indent Explicit analytical results are given in the Laplace
domain -the zeroth order by Eq.~(\ref{eq50}) and higher order
terms by Eqs.~(\ref{eq52}) and (\ref{eq54}); then numerical
inversion is required to transform them into the temporal domain.
In detail, we performed a standard numerical integration (function
NIntegrate in the Mathematica package) between proper limits in
the imaginary axis, according to the definition of the inverse
Laplace transform with a real integration variable,

\begin{equation}\label{eq69}
   f(\tau) = \frac{1}{2\pi} ~ \int_{-{\rm j}\omega}^{+{\rm
   j}\omega} \tilde{f}^{L}({\rm j}\omega) ~ \rme^{{\rm j}\omega\tau}~
   \rmd \omega,
\end{equation}

\noindent with $\omega$ properly chosen to represent infinity.
According to integration parameters and the function being inverse
transformed, numerical instabilities not associated with the
validity of the series method itself may arise. However, in all
the cases presented here, the analytical results in the temporal
domain were consistent for different large $\omega$ limits and
numerical parameters. In figure \ref{fig1}, the FPT statistics in
the sub/supra-threshold regimes, with low/high noise intensities,
and different strengths of the time-dependent drift are shown. In
all cases, the time constant of the exponential drift $\tau_{\rm
d}$ is set to $100$ time units (ms). As observed, the series
solution up to a certain order $N$, represented by the thin yellow
lines, properly describes the FPT distributions obtained from the
numerical simulations (stair-like colored histograms). However,
different parameter combinations require different $N$ to
converge. In general, when noise intensity is reduced, the FPT
distribution becomes more sharply peaked (or distorted) in
comparison to the unperturbed case, and the order of convergence
increases concomitantly. This can be noticed in figures
\ref{fig1}{\it(a)} and \ref{fig1}{\it(c)}, where the noise
intensity has been manipulated in a generic sub-threshold regime,
or in figures \ref{fig1}{\it(b)} and \ref{fig1}{\it(d)} in a
general supra-threshold condition. Naturally, larger amplitudes of
the time-dependent drift require more terms to be considered in
the series expansion to converge to a final distribution,
supposing the parameters are within the radius of convergence.
This increase in the number of terms may affect convergence in the
temporal domain -regardless of the radius of convergence of the
series itself- because of the amplification of numerical errors
and/or instabilities when a specific procedure with a given set of
numerical parameters is used to inverse Laplace transform
functions.

\begin{figure}[t!]
\begin{center}
\includegraphics[scale=1.0]{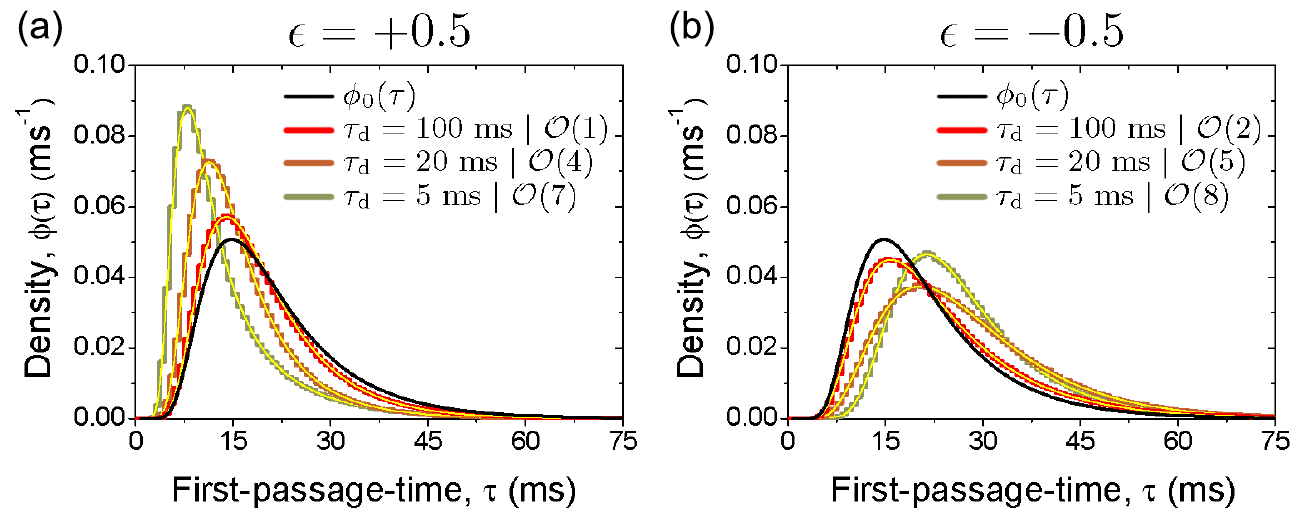}
\caption{\label{fig2} Comparison between the analytical and
numerical results of the FPT statistics for different timescales
defining time-dependent drift. Relatively moderate positive,
$\epsilon = +0.5$, and negative, $\epsilon = -0.5$, intensities
are considered in (a) and (b), respectively. FPT distributions for
different time constants $\tau_{\rm d}$, obtained from numerical
simulations are shown by different stair-like colored lines.
Analytical results up to the order $N$ indicated in the
upper-right hand part of each panel, $\mathcal{O}(N)$, are
represented by thin yellow lines. As observed, the agreement
between the numerical results and theoretical expressions is
excellent in all cases, but the order of convergence may differ
substantially. Parameters governing the dynamics are $x_0 = 0$,
$x_{\rm thr} = 1$, $\tau_{\rm m} = 10~{\rm ms}$, $\mu = 0.100~{\rm
ms}^{-1}$, and $D = 0.005~{\rm ms}^{-1}$.}
\end{center}
\end{figure}

\indent As the time constant of the time-dependent exponential
drift decreases, the series solution requires more higher order
terms to become accurate. This general behavior is depicted in
figure \ref{fig2}, where the theoretical series solution (thin
yellow lines) correctly represents FPT distributions obtained from
numerical simulations (stair-like colored histograms), for
positive as well as negative intensities $\epsilon$ (figure
\ref{fig2}{\it(a)} and \ref{fig2}{\it(b)}, respectively) for
different time constants $\tau_{\rm d}$. According to the general
expression for the higher order terms in the series solution,
Eqs.~(\ref{eq52}) and (\ref{eq54}), the coefficients weighting
individual contributions for a given term, Eqs.~(\ref{eq55}) and
(\ref{eq56}), are extremely sensitive to the ratio between the
extrinsic and intrinsic timescales, $\tau_{\rm d}/\tau_{\rm m}$.
Given this dependence (and assuming that convergence exists), the
order of convergence of the series solution should increase as
$\tau_{\rm d}/\tau_{\rm m} \rightarrow 1$. In the examples
analyzed here, convergent behavior was observed for consecutive
terms in the series solution as this limit was reached. The
special case $\tau_{\rm m} = \tau_{\rm d}$ can not be described by
the previous formulae, and a Green's function approach has been
taken (see Appendix). On the other hand, as the time constant
approaches zero, $\tau_{\rm d} \rightarrow 0$, the superimposed
time-dependent exponential drift diverges and the system
effectively corresponds to an unperturbed case with a shift in the
initial condition \cite{Urdapilleta2011a}. In this limit,
according to the value of $\epsilon$, the FTP problem will be
well-posed only when this initial condition is below the
threshold.

\indent The series solution for the FPT density function,
Eq.~(\ref{eq32}), is explicitly given in terms of its Laplace
transform: the zeroth order by Eq.~(\ref{eq50}), and higher order
terms by Eqs.~(\ref{eq52}) and (\ref{eq54}). Without performing
any inverse transformation (and, therefore, avoiding all numerical
artifacts of numerical implementation), this result can be used to
obtain important properties of the FPT distribution. In
particular, its moments are given by \cite{Urdapilleta2011a,
Urdapilleta2012}

\begin{equation}\label{eq70}
   \langle \tau^{k} \rangle = \int_{0}^{\infty}
   \phi(\tau)~\tau^{k}~{\rm d}\tau = (-1)^{k}~\frac{{\rm d}^{k} \tilde{\phi}^{L}(s)}{{\rm
   d}s^{k}}\Big\rfloor_{s=0}.
\end{equation}

\indent Since the series solution for $\phi(\tau)$ is a linear
combination of different functions, the preceding relationship can
be written as

\begin{equation}\label{eq71}
   \langle \tau^{k} \rangle = \sum_{n=0}^{\infty} \epsilon^{n}~\langle \tau^{k}
   \rangle_{\phi_{n}},
\end{equation}

\noindent where

\begin{equation}\label{eq72}
   \langle \tau^{k} \rangle_{\phi_{n}} = (-1)^{k}~\frac{{\rm d}^{k} \tilde{\phi}_{n}^{L}(s)}{{\rm
   d}s^{k}}\Big\rfloor_{s=0}.
\end{equation}

\begin{figure}[t!]
\begin{center}
\includegraphics[scale=1.0]{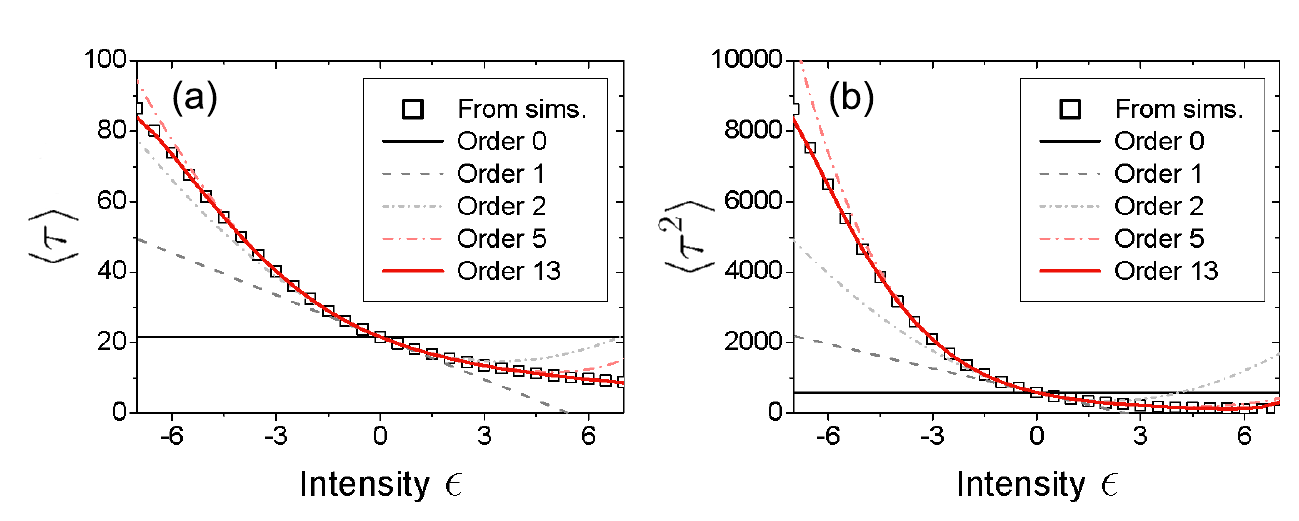}
\caption{\label{fig3} Comparison between analytical and numerical
results for (a) the first and (b) the second moment of the FPT
distribution, as a function of the intensity of time-dependent
drift $\epsilon$. The symbols represent the average of the
numerical results, whereas the lines correspond to the series
expression for the $k$-th moment truncated at the order $N$
indicated in the inset, $\langle \tau^{k} \rangle = \sum_{n =
0}^{N} \epsilon^{n} ~ \langle \tau^{k} \rangle_{\phi_{n}} $. The
remaining parameters are $x_0 = 0$, $x_{\rm thr} = 1$, $\tau_{\rm
m} = 10~{\rm ms}$, $\tau_{\rm d} = 100~{\rm ms}$, $\mu =
0.100~{\rm ms}^{-1}$, and $D = 0.005~{\rm ms}^{-1}$.}
\end{center}
\end{figure}

\indent According to Eqs.~(\ref{eq71}) and (\ref{eq72}), the
evaluation of the moments is straightforward in the Laplace
domain. In figure \ref{fig3} we show the behavior of the first two
moments for the intermediate case analyzed in figure \ref{fig2},
as a function of a wide range of the intensity $\epsilon$. As
observed, as the absolute value of $\epsilon$ increases, the order
$N$ considered in the series has to increase as well to account
for the numerical results.

\section{Final remarks and discussion}\label{discussion}
\indent In this work, we have studied the survival probability and
the FPT statistics of a Brownian particle whose dynamics are
governed by a generic unidimensional potential and a superimposed
exponential time-dependent drift, Eq.~(\ref{eq10}), with a fixed
threshold setting the limit where the FPT is defined. Based on the
backward FP description, we first derived the diffusion equation
of the survival probability from the backward state,
Eq.~(\ref{eq12}), and then proposed a series solution in powers of
the intensity of the time-dependent drift contribution,
Eq.~(\ref{eq13}). With this procedure, the preceding diffusion
equation translates into a system of infinite simpler equations,
where each one defines the behavior of each term in the proposed
series in a recursive scheme, Eqs.~(\ref{eq15}) and (\ref{eq16}).
The particular mathematical structure of the forcing term in these
equations -defined by the time-dependent exponential drift-
enables a simpler representation in the Laplace domain,
Eqs.~(\ref{eq22}) and (\ref{eq30}). From the survival probability,
the FPT statistics are readily obtained, which naturally inherit
the series structure, Eq.~(\ref{eq32}) where each term is governed
by Eqs.~(\ref{eq36}) or (\ref{eq37}). The general derivation of
this series solution agrees with the explicit solution we found
previously for the Wiener process \cite{Urdapilleta2012}. However,
since the present approach is applicable to any unidimensional
potential, we explicitly solved the series solution for the FPT
statistics of an Ornstein-Uhlenbeck process (with the superimposed
exponential time-dependent field), which is mathematically much
more challenging than the Wiener process. Given that the
convergence properties of the proposed series solution remained
unknown, several cases were defined to numerically test the
usefulness of the approach and the solution found. In all cases,
the analytical and numerical results were in good agreement as
long as the number of terms included in the series is large
enough.

\indent As discussed in Section \ref{intro}, different neuron
models have a direct correspondence to different drift-diffusion
Brownian motions. In particular, the Wiener and the
Ornstein-Uhlenbeck processes correspond to the perfect and leaky
IF neuron models, respectively. In all IF models, variable $x$ is
the transmembrane voltage of a spiking neuron, the drift derived
from the potential, $-U'(x)$, corresponds to the external as well
as the internal sub-threshold signals integrated by the Langevin
dynamics, Eq. ~(\ref{eq1}), which model the capacitance properties
of the cellular membrane. Additive Gaussian white noise is
included in order to model randomness arising from different
sources (random channel opening and closing, stochastic synaptic
transmission, etcetera), with minimal mathematical complexity. The
FPT problem results from the procedure used to declare a spike.
Whenever the voltage reaches the threshold $x_{\rm thr}$, the
voltage dynamics are no longer governed by the simple Langevin
equation, Eq.~(\ref{eq1}), and a largely stereotyped waveform is
generated by other mechanisms (not included in IF models).
Therefore, to a large extent, the exact time at which this event
was produced \textit{for the first time} is the only meaningful
information to be transmitted to downstream neurons. Once this
event happens, there are mechanisms that restore the membrane
potential to a reset potential $x_0$ (here also coincident to the
resting potential), and a new sub-threshold integration cycle is
launched. The superimposed time-dependent exponential drift
considered here mimics the temporal evolution of a separate
process of neuronal adaptation, which naturally also influences
spike time statistics. This adaptation current not only modifies
the FPT statistics of the homogeneous case (for example, the pure
Wiener or Ornstein-Uhlenbeck processes or, equivalently, the
adaptation-free perfect and leaky IF models), but also couples
subsequent events, creating negative correlations
\cite{Urdapilleta2011b, SchwalgerLindner2013}. The analysis of
successive interspike intervals and the emergence of these
correlations can be described by a hidden Markov model
\cite{Urdapilleta2011b}, in which correlations arise from the
fire-and-reset rule coupling the subsequent initial state of the
adaptation current with the preceding interspike interval. In this
description, the FPT statistics of the temporally inhomogeneous
process considered here provide the relationship between the
hidden variable (the initial state of the adaptation current) and
the observable (the interspike intervals).

\indent Since both Wiener and Ornstein-Uhlenbeck processes as well
as generic drift-diffusion models are ubiquitous in describing
different phenomena, the results obtained here on survival
probability and FPT distribution will be of interest in other
settings where an additive state-independent temporal relaxation
process is being developed as the particle diffuses.

\section{Acknowledgments}
This work was supported by the Consejo de Investigaciones
Cient\'ificas y T\'ecnicas de la Rep\'ublica Argentina.

\section*{Appendix}
\subsection*{FPT statistics for the Ornstein-Uhlenbeck process, with $\tau_{\rm m}
= \tau_{\rmd}$}

\indent Naturally, the equation governing the behavior of the
unperturbed system, Eq.~(\ref{eq36}), does not depend on the
timescale of the exponential time-dependent drift, $\tau_{\rmd}$.
On the other hand, when $\tau_{\rm m} = \tau_{\rmd}$,
Eq.~(\ref{eq37}) applied to the Ornstein-Uhlenbeck case,
Eq.~(\ref{eq51}), reads

\begin{equation}\label{ap1}
   \fl \hspace{1.cm}
   \frac{\rmd^{2}\tilde{\phi}_{n}^{L}}{\rmd x_0^2}
   + \left( \frac{\mu}{D} - \frac{x_0}{\tau_{\rm m} D} \right) ~
   \frac{\rmd\tilde{\phi}_{n}^{L}}{\rmd x_0}
   - \frac{(s+n/\tau_{\rm m})}{D}~\tilde{\phi}_{n}^{L}
   = -\frac{1}{\tau_{\rm m}D} ~ \frac{\rmd \tilde{\phi}_{n-1}^{L}}
   {\rmd x_0},  ~n \geq 1.
\end{equation}

\indent Again, with the substitution $z = \sqrt{\tau_{\rm m}/D}~(
\mu - x_0/\tau_{\rm m})$ and proposing the functional structure
given by Eq.~(\ref{eq52}), the preceding equation transforms to

\begin{equation}\label{ap2}
   \fl \hspace{1.cm}
   \frac{\rmd^{2}u_{n}}{\rmd z^2} +
   \left[ -\tau_{\rm m}\left( s + \frac{n}{\tau_{\rm m}} \right)
   + \frac{1}{2} - \frac{1}{4} z^2 \right]~u_{n} = \frac{1}
   {\sqrt{\tau_{\rm m}D}}~\left(
   \frac{1}{2}~z~u_{n-1} + \frac{\rmd u_{n-1}}{\rmd z} \right).
\end{equation}

\indent As before, boundary conditions are $u_{n}(z_{\rm thr}) =
0$ and $u_{n}(z \rightarrow \infty)$ bounded, where the dependence
on the parameter $s$ has been omitted for simplicity. The solution
to this infinite set of equations can be recursively found as

\begin{equation}\label{ap3}
   u_{n}(z) = \int_{z_{\rm thr}}^{\infty} g_{n}(z,z')~\frac{1}
   {\sqrt{\tau_{\rm m}D}}~\left[ \frac{1}{2}~z'~u_{n-1}(z') +
   \frac{\rmd u_{n-1}(z')}{\rmd z'} \right]~{\rmd}z',
\end{equation}

\noindent where $g_{n}(z,z')$ is the Green's function of
Eq.~(\ref{ap2}), which is obtained as the solution to

\begin{equation}\label{ap4}
   \frac{\partial^{2}g_{n}(z,z')}{\partial z^2} +
   \left[ -\tau_{\rm m}\left( s + \frac{n}{\tau_{\rm m}} \right)
   + \frac{1}{2} - \frac{1}{4} z^2 \right]~g_{n}(z,z') =
   \delta(z-z'),
\end{equation}

\noindent with the boundary conditions defined above. Explicitly,
this function reads

\begin{equation}\label{ap5}
   \fl \hspace{1.cm}
   g_{n}(z,z') = \frac{1}{{\rm Den}(z')} \left\{
   \eqalign{
      \Big[ \mathcal{D}_{\nu}(z_{\rm thr})~\mathcal{D}_{\nu}(-z) &
      - \mathcal{D}_{\nu}(-z_{\rm thr})~\mathcal{D}_{\nu}(z) \Big]
      ~\mathcal{D}_{\nu}(z') ~,~{\rm for}~~ z<z', \cr
      \Big[ \mathcal{D}_{\nu}(z_{\rm thr})~\mathcal{D}_{\nu}(-z') &
      - \mathcal{D}_{\nu}(-z_{\rm thr})~\mathcal{D}_{\nu}(z') \Big]
      ~\mathcal{D}_{\nu}(z) ~,~{\rm for}~~ z>z',
   }
   \right.
\end{equation}

\noindent where

\begin{equation}\label{ap5}
   \fl \hspace{1.cm}
   {\rm Den}(z') = \frac{1}{\nu~\mathcal{D}_{\nu}(z_{\rm thr})~
   \Big[ \mathcal{D}_{\nu}(-z')~\mathcal{D}_{\nu-1}(z')
   + \mathcal{D}_{\nu}(z')~\mathcal{D}_{\nu-1}(-z') \Big]},
\end{equation}

\noindent and the order $n$ of the function $g_n(z,z')$ comes into
the index $\nu = -\tau_{\rm m} \left(s+n/\tau_{\rm m}\right)$
exclusively.

\end{document}